\documentclass[runningheads]{llncs}

\usepackage{eccv}

\usepackage{eccvabbrv}

\usepackage{graphicx}
\usepackage{booktabs}
\usepackage{multirow}
\usepackage{marvosym}
\usepackage{makecell}

\usepackage[accsupp]{axessibility}  

\newcommand{\equalcontrib}{\textsuperscript{*}}
\newcommand{\corrauth}{\textsuperscript{\textrm{\Letter}}}

\usepackage{hyperref}

\usepackage{orcidlink}

\begin{document}

\title{MedGSSR: Generalizable Medical Image Super-Resolution 3D Reconstruction via Hierarchical Feed-forward Gaussian Splatting} 

\titlerunning{MedGSSR}

\author{
Chengkai Wang\inst{1}\equalcontrib \and
Luoyu Hong\inst{2}\equalcontrib \and
Yiting Zhao\inst{2} \and
Jiamin Wang\inst{3} \and
Xiang Feng\inst{3} \and
Feiwei Qin\inst{2} \and
Zhenzhong Kuang\inst{2} \and
Xuefei Yin\inst{4} \\
Ali Bashashati\inst{1} \corrauth \and
Yanming Zhu\inst{4}\corrauth
}

\authorrunning{Wang et al.}

\institute{University of British Columbia, Vancouver, Canada
 \and Hangzhou Dianzi University, Hangzhou, China  \and ShanghaiTech University, Shanghai, China \and Griffith University, Gold Coast, Australia
}

\maketitle

\begingroup
\renewcommand{\thefootnote}{\fnsymbol{footnote}}
\footnotetext[1]{Equal contribution. \quad \corrauth\ Corresponding authors.}
\endgroup

\vspace{-1.5em}

\begin{abstract}
High-resolution volumetric medical imaging is critical for clinical diagnosis, yet acquisition is often limited by scanner hardware, scan time, and for CT, radiation dose. Medical 3D Super-Resolution (Med3DSR) offers a computational alternative, but existing methods commonly rely on per-subject optimization, pretrained priors, or coordinate-based implicit representations, which compromise anatomical fidelity and limit efficiency. To address these limitations, we present \textbf{MedGSSR}, a fully end-to-end feed-forward framework that represents volumes as an explicit 3D Gaussian field for Med3DSR. Unlike coordinate-based implicit functions, our explicit 3D Gaussian representation naturally enhances signal continuity and local high-frequency fidelity. Specifically, MedGSSR explicitly decouples the reconstruction process into coarse-grained structural preservation and fine-grained textural refinement through the proposed Pyramid Anatomical Encoder and a Hierarchical Gaussian Projector. To support arbitrary-scale super-resolution, we introduce sub-voxel Gaussian decomposition and a Differentiable Gaussian Voxelizer that directly queries the continuous 3D intensity field, reducing discretization artifacts. Extensive experiments on MRI and CT benchmarks demonstrate that MedGSSR significantly outperforms state-of-the-art methods. Notably, our framework exhibits robust generalizability across unseen datasets without requiring per-subject optimization, enabling fast inference and high-fidelity volumetric super-resolution in practical clinical settings. Our project webpage,
including code, is at \url{https://william2ai.github.io/medgssr}

  \keywords{3D Medical Image Super-Resolution \and  Feed-forward Gaussian Splatting \and Arbitrary-Scale Reconstruction } 
\end{abstract}

\section{Introduction}

High-resolution (HR) volumetric medical imaging, such as MRI and CT, is fundamental to accurate clinical diagnosis and quantitative analysis. However, obtaining isotropic HR volumes is often constrained by physical and practical limitations. High-field MRI scanners remain scarce, and prolonged scanning times required for high spatial resolution increase the risk of motion artifacts \cite{forigua2022superformer, wang2025mmdental}. Similarly, in CT imaging, increasing resolution inevitably requires higher radiation doses, heightening safety risks \cite{krug2010high}. To bridge the gap between the observed low-resolution (LR) data and the clinically desired HR volumes, Med3DSR has emerged as a critical research focus, aiming to computationally recover missing spatial information without modifying the imaging hardware.

\begin{figure*}[t]
    \centering
    \includegraphics[width=\textwidth]{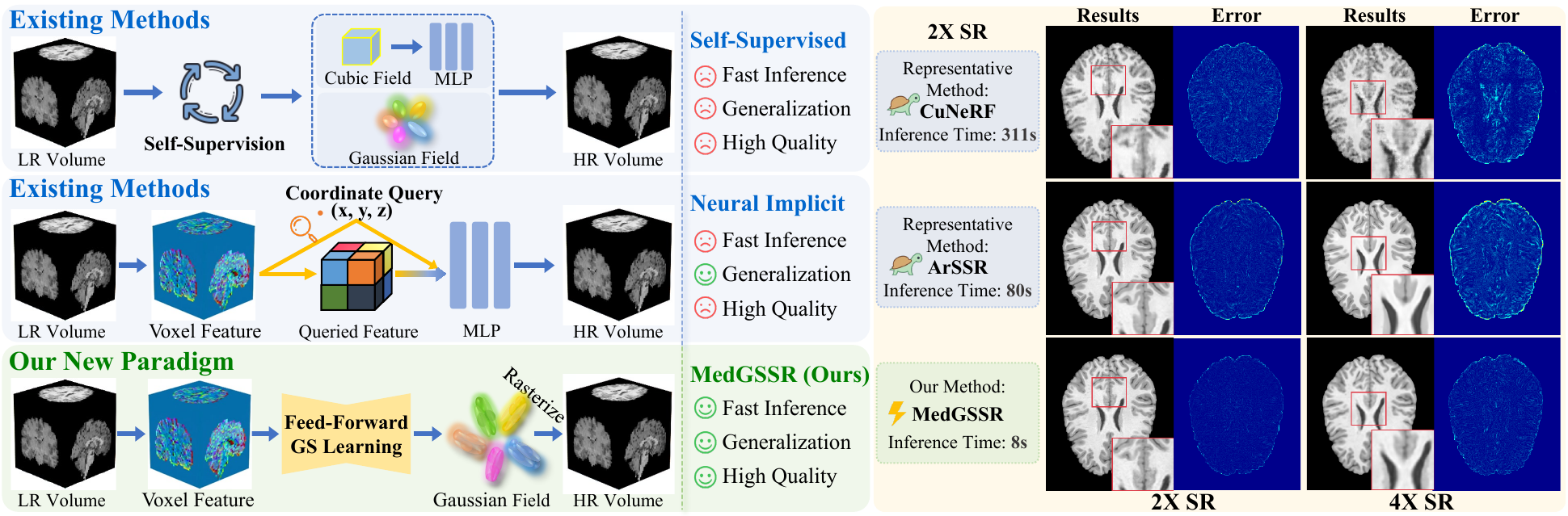}
    \caption{MedGSSR introduces a paradigm shift for Medical 3D Super-Resolution. Left: Existing self-supervised methods (relying on isolated optimization) and neural implicit networks (relying on coordinate querying) fail to simultaneously achieve fast inference, robust generalization, and high-quality reconstruction. MedGSSR introduces a generalized feed-forward 3DGS paradigm that explicitly satisfies all three clinical requirements. Right: Cross-dataset Visual comparisons (trained on MSD, tested on HCP) at $2\times$ and $4\times$ upsampling scales. Compared to baselines that produce geometric artifacts or over-smoothed textures, MedGSSR accurately restores high-frequency anatomical structures with the lowest error (dark blue in difference maps) while accelerating inference by an order of magnitude.}
    \label{fig:teaser}
\end{figure*}

Despite recent progress, clinical-grade Med3DSR remains challenged by two persistent bottlenecks, as shown in Fig.~\ref{fig:teaser}. First, continuous zooming does not necessarily translate to high-frequency fidelity. Implicit Neural Representations (INRs) enable efficient, arbitrary-scale querying by learning a coordinate-to-intensity mapping \cite{wu2022arbitrary}. However, coordinate-based implicit functions face intrinsic representational limitations. Because a single set of shared network weights must represent the entire volumetric signal, coordinate-based INRs are prone to spectral bias \cite{rahaman2019spectral}, often fitting low-frequency components and producing over-smoothed reconstructions that miss fine anatomical structures and pathological boundaries.
Second, limited isotropic 3D data hinders learning 3D-consistent priors and generalization. Due to the scarcity of high-quality isotropic 3D medical data, explicit reconstruction pipelines retreat to either isolated per-subject optimization \cite{chen2023cunerf} or 2D proxy supervision \cite{noh2025zero}. Per-subject optimization treats each scan as an independent fitting problem, which is computationally prohibitive and prevents transferable anatomical priors from being learned. Lacking holistic 3D spatial context, these slice-wise 2D proxies often introduce inter-slice inconsistencies \cite{hoeg2024mtvnet} and texture artifacts, making the attainable 3D reconstruction fidelity strongly dependent on the quality of the 2D pseudo-labels \cite{feng2024zs, feng2026ie}.

To address these limitations, we propose MedGSSR, a novel paradigm that reformulates Med3DSR as a generalized, purely volumetric feed-forward prediction of an explicit 3D Gaussian Splatting (3DGS) field. 
In contrast to coordinate-based INRs that encode the entire volume in shared global weights, MedGSSR represents anatomy as a set of local-support Gaussian primitives, which reduces global coupling across distant regions and facilitates localized modeling of high-frequency structures. 
Furthermore, since medical modalities directly measure continuous physical density (e.g., Hounsfield units in CT or proton density in MRI), modeling the anatomy as an explicit superposition of continuous Gaussian distributions ensures strict physical consistency. 
Moreover, unlike optimization-based methods that isolate each scene, MedGSSR learns a direct feed-forward mapping function from sparse LR voxel grids to continuous HR 3DGS parameters. By training on large-scale multi-subject datasets, our model autonomously learns 3D-specific anatomical priors, reconstructing accurate high-frequency details via continuous Gaussian representations across unseen subjects without test-time optimization.

To effectively adapt this paradigm for highly heterogeneous medical data, we design a specialized architecture. We introduce a Pyramid Anatomical Encoder that extracts multi-scale volumetric features and organizes them into two complementary representations capturing global anatomy and local appearance. Subsequently, a Hierarchical Gaussian Projector explicitly factorizes reconstruction into two branches: a coarse branch dedicated to preserving macroscopic organ geometry, and a fine branch focused on refining local tissue textures. To bridge the gap between discrete feature grids and continuous biological signals, we employ a sub-voxel decomposition strategy, where each voxel is represented by a set of adaptive Gaussian primitives, effectively eliminating discretization artifacts. Finally, a Differentiable Gaussian Voxelizer aggregates these primitives into a continuous 3D volumetric field. Consequently, this formulation allows us to render HR volumes at any desired resolution by simply querying the intensity field at corresponding spatial coordinates, entirely within the 3D domain.

The core contributions of our study are as follows:
\begin{itemize}
    \item We reformulate Med3DSR as a generalized, purely volumetric feed-forward mapping from sparse LR voxel grids to an explicit continuous 3DGS field, shifting the paradigm from INRs and per-subject optimization to a generalizable physics-consistent explicit representation that enables fast inference, high-fidelity reconstruction, and cross-subject generalization.
    \item  We propose a Hierarchical Gaussian Projector with a sub-voxel decomposition mechanism, which explicitly decouples structural preservation from textural refinement to handle the strong heterogeneity of medical images.
    \item Extensive experiments on MRI and CT benchmarks demonstrate that the proposed MedGSSR outperforms SOTA approaches, achieving superior high-frequency reconstruction fidelity and robust cross-dataset generalization.
\end{itemize}

\section{Related Work}

\subsection{3D Reconstruction}
The evolution of recent neural 3D scene reconstruction has been heavily driven by Implicit Neural Representations, notably NeRF \cite{mildenhall2021nerf}, which models continuous volumetric fields using MLPs. Subsequent variants improved rendering quality and optimization speed by integrating multi-scale representations \cite{barron2021mip}, hash-grid encodings \cite{muller2022instant}, and tensorial decomposition \cite{chen2022tensorf}. To overcome the computational bottleneck of volume rendering, 3DGS \cite{kerbl20233d} introduced an explicit point-based formulation for efficient rasterization, with methods like Mip-Splatting \cite{yu2024mip} further addressing aliasing artifacts.

In the domain of 3D super-resolution (3DSR), early implicit methods \cite{wang2022nerf, yoon2023cross} and recent 3DGS-based models \cite{feng2024srgs, xie2024supergs} rely heavily on pretrained 2D single-image super-resolution (SISR) networks to generate pseudo-HR labels. While improving visual quality, these methods require time-consuming per-subject optimization. Furthermore, relying on 2D models to process individual viewpoints independently fundamentally lacks 3D spatial awareness. This mechanism inevitably introduces cross-view inconsistencies and caps reconstruction fidelity at the upper bound of the 2D pseudo-labels.

To bypass per-subject optimization, recent vision models \cite{charatan2024pixelsplat, szymanowicz2024splatter, xu2025depthsplat, ye2024no} have explored the direct feed-forward prediction of 3DGS parameters from sparse images. However, these architectures target sparse natural scenes and are structurally ill-equipped for the dense, continuous topologies of volumetric medical data. In contrast to these approaches, our method explicitly adapts the feed-forward 3DGS paradigm for medical volumes to establish generalized, 3D-consistent structural priors.

\subsection{Medical Image Super-Resolution}

Early data-driven medical image super-resolution methods predominantly relied on 3D CNNs, including SRCNN3D \cite{7950500}, mDCSRN \cite{chen2018efficient}, and ResCNN \cite{du2020super}. While these architectures effectively restore local textures compared to traditional interpolation, their heavy computational demands and restriction to fixed integer upsampling scales limit their clinical flexibility.

To achieve arbitrary-scale super-resolution, the focus shifted toward INRs \cite{wu2022arbitrary, chen2023cunerf, fang2024cycleinr}, which map continuous spatial coordinates to intensity values. Additionally, self-supervised learning frameworks \cite{shocher2018zero} train exclusively on the input volume to avoid the need for external HR data. However, coordinate-based MLPs inherently suffer from spectral bias, prioritizing low-frequency structures and producing over-smoothed textures. Concurrently, self-supervised methods fail to leverage large-scale anatomical priors due to their isolated training mechanisms.

Recently, researchers have attempted to apply explicit 3D representations to medical super-resolution. Due to the scarcity of paired 3D HR data, methods such as MedNeRF \cite{corona2022mednerf} and diffusion-guided NAB-GS \cite{noh2025zero} project volumes into 2D X-ray domains to exploit 2D generative priors. This 3D-to-2D-to-3D pipeline inevitably introduces modality gaps and geometric ambiguity during projection. Unlike these projection-based or implicit approaches, MedGSSR operates purely in the continuous 3D domain, avoiding spectral bias and modality gaps entirely.

\section{Preliminaries}
In this section, we briefly review the standard formulation of 3DGS and discuss the paradigm shift toward feed-forward prediction, which forms the theoretical foundation of our MedGSSR framework.
\subsection{3D Gaussian Splatting (3DGS)}
\label{sec:prelim_3dgs}

3DGS models a volumetric scene using a collection of anisotropic 3D Gaussian primitives $\mathcal{G} = \{G_1, \dots, G_N\}$. In contrast to INRs that implicitly encode scene properties within neural network weights, 3DGS explicitly parameterizes geometry and appearance. Each primitive $G_i$ is characterized by a center position $\boldsymbol{\mu}_i \in \mathbb{R}^3$, a covariance matrix $\boldsymbol{\Sigma}_i \in \mathbb{R}^{3\times3}$, an opacity scalar $\xi_i \in[0, 1]$, and appearance features $\boldsymbol{c}_i$. The spatial contribution of the $i$-th Gaussian at a query point $\boldsymbol{x} \in \mathbb{R}^3$ is given by:
\begin{equation}
    G_i(\boldsymbol{x}) = \xi_i \cdot \exp \left( -\frac{1}{2} (\boldsymbol{x} - \boldsymbol{\mu}_i)^\top \boldsymbol{\Sigma}_i^{-1} (\boldsymbol{x} - \boldsymbol{\mu}_i) \right).
    \label{eq:gaussian_def}
\end{equation}
These Gaussians are projected into screen space and rasterized via alpha blending to render novel views. To ensure $\boldsymbol{\Sigma}_i$ is symmetric positive definite (and thus invertible), it is typically parameterized as a scaling vector $\boldsymbol{s}_i$ and a rotation quaternion $\boldsymbol{q}_i$. Consequently, the full learnable parameter set becomes $\Theta = \{\boldsymbol{\mu}_i, \boldsymbol{q}_i, \boldsymbol{s}_i, \xi_i, \boldsymbol{c}_i\}$, which is optimized via gradient descent to minimize the reconstruction error between rendered projections and ground-truth images.

\subsection{Feed-Forward 3D Gaussian Prediction}
\label{sec:prelim_feedforward}

In contrast to standard 3DGS which entails computationally intensive per-subject optimization and lacks generalization across subjects, recent advances in computer vision have recast scene reconstruction as a direct regression task. While existing feed-forward models primarily map sparse 2D images to 3D scenes, we reconceptualize this paradigm for continuous volumetric medical data. 

Specifically, we formulate the Med3DSR task as a direct voxel-to-Gaussian mapping. Given a low-resolution input observation $\mathbf{V}_{LR}$ (i.e., a 3D voxel grid), our objective is to learn a mapping function $\Psi_\theta$, parameterized by a neural network, that directly predicts the Gaussian attributes for any given spatial location $\boldsymbol{x}$:
\begin{equation}
    \{\boldsymbol{\mu}, \boldsymbol{q}, \boldsymbol{s}, \xi, \boldsymbol{c}\}_{\boldsymbol{x}} = \Psi_\theta(E(\mathbf{V}_{LR}))_{\boldsymbol{x}},
\end{equation}
where $E(\cdot)$ denotes a volumetric feature extraction encoder. This explicit formulation allows the network to predict a dense, continuous field of Gaussian primitives directly from 3D medical scans.

\begin{figure*}[t]
    \centering
    \includegraphics[width=\textwidth]{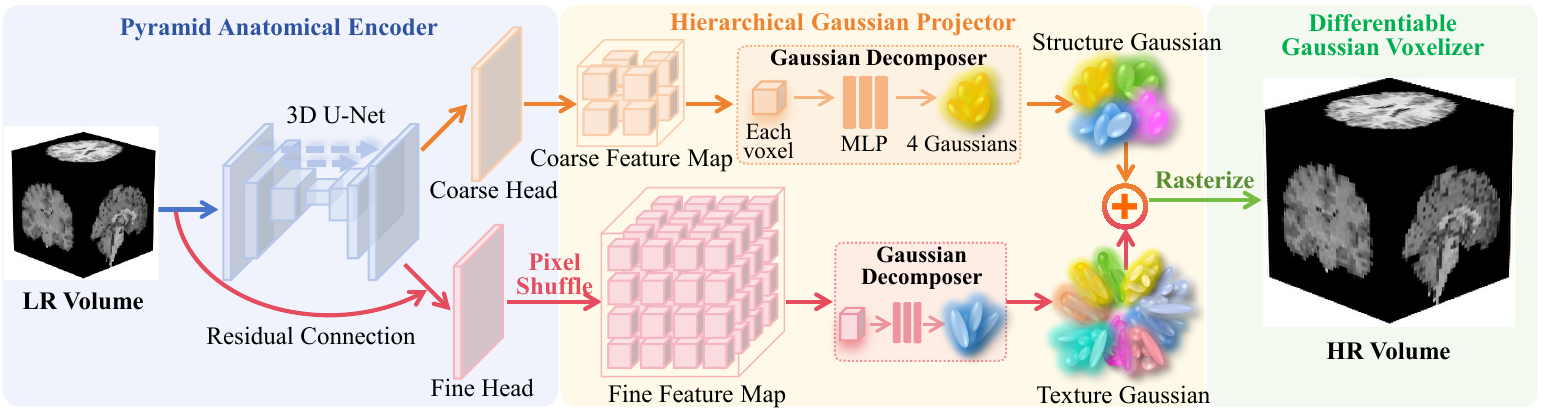}
    \vspace{-0.2cm}
    \caption{Overview of MedGSSR. Given an LR input volume, the Pyramid Anatomical Encoder (PAE) built on a 3D U-Net extracts a coarse feature map and a $2\times$ upsampled fine feature map via input-residual connection and PixelShuffle. The Hierarchical Gaussian Projector (HGP) performs sub-voxel Gaussian decomposition by predicting multiple Gaussian primitives per voxel with lightweight MLP heads, producing structural Gaussians from the coarse stream and textural Gaussians from the fine stream. These primitives are then aggregated by the Differentiable Gaussian Voxelizer (DGV) through differentiable rasterization to form a continuous volumetric intensity field, from which the HR volume is rendered.}
    \label{fig:overview}
\end{figure*}

\section{Methodology}

\subsection{Overview}
Given a low-resolution (LR) medical volumetric input $\mathbf{V}_{LR} \in \mathbb{R}^{H \times W \times D}$, our goal is to reconstruct a high-fidelity super-resolved volume $\mathbf{V}_{HR}$ at an arbitrary upsampling scale $\boldsymbol{\kappa}$. MedGSSR reformulates this task as a direct prediction problem, learning a generalizable mapping function $\Psi: \mathbf{V}_{LR} \to \mathcal{G}$, where $\mathcal{G}$ represents a continuous explicit field of 3D Gaussian primitives, as illustrated in Fig.~\ref{fig:overview}.

\subsection{Pyramid Anatomical Encoder}
To effectively capture the heterogeneous nature of medical scans, where macroscopic organ geometry and microscopic tissue textures coexist, we design the PAE as a dual-stream feature extractor built upon a 3D U-Net architecture. We deliberately eschew Batch Normalization layers throughout the network to preserve absolute intensity information (e.g., Hounsfield Units in CT), which is clinically critical for accurate diagnosis.
Specifically, the 3D U-Net processes $\mathbf{V}_{LR}$ and outputs the final decoder feature map $\mathbf{H}_{dec}$. The coarse stream then maps $\mathbf{H}_{dec}$ to a coarse feature map $\mathbf{F}_{coarse}\in\mathbb{R}^{H \times W \times D \times C}$ via a 3D convolution layer. Defined on the original $1\times$ voxel grid, $\mathbf{F}_{coarse}$ preserves spatial alignment while capturing organ-scale context. Compared with $\mathbf{F}_{fine}$ defined below, each voxel in $\mathbf{F}_{coarse}$ corresponds to a $2\times$ larger spacing along each spatial axis on the same underlying volume, and thus covers an $8\times$ larger physical volume. This coarse representation therefore provides a large receptive field for capturing global anatomical structures.

For the fine-grained stream, to enhance high-frequency details while retaining structural cues, we introduce an \emph{input-residual connection} through concatenation. Specifically, we concatenate the LR input $\mathbf{V}_{LR}$ as a residual signal with the decoder feature map $\mathbf{H}_{dec}$ along the channel dimension. The concatenated tensor is then passed through a 3D convolution layer to expand channels, followed by a 3D PixelShuffle operator with an upsampling factor of $2$:
\begin{equation}
    \mathbf{F}_{fine} = \text{PixelShuffle}(\text{Conv}_{3D}([\mathbf{H}_{dec}, \mathbf{V}_{LR}])) \in \mathbb{R}^{2H \times 2W \times 2D \times C'},
\end{equation}
where $[\cdot,\cdot]$ denotes channel-wise concatenation. This dual-resolution design provides complementary coarse-to-fine features for subsequent Gaussian projection.

\subsection{Hierarchical Gaussian Projector (HGP)}

\textbf{Structure-Texture Decoupling.} The HGP comprises two parallel projection heads
implemented as dedicated MLPs to directly regress Gaussian attributes. The structure-texture decoupling is fundamentally enabled by our dual-resolution feature design.
The coarse head operates on the $1\times$ resolution feature map $\mathbf{F}_{coarse}$, where each voxel corresponds to a larger physical neighborhood on the underlying volume and thus emphasizes organ-scale, low-frequency anatomical context. Conversely, the fine head operates on the $2\times$ upsampled feature map $\mathbf{F}_{fine}$, which provides denser spatial sampling and richer local cues, and is tailored to refine microscopic, high-frequency details and boundaries.

\textbf{Sub-voxel Gaussian Decomposition.} A core challenge in volumetric SR is the discretization artifact arising from sparse voxel grids, which limits the ability to represent sub-voxel intensity variations. To address this, we propose a Sub-voxel Gaussian Decomposition strategy within the HGP. Instead of assigning a single Gaussian primitive to each voxel location, we predict a set of $m$ Gaussian primitives with learnable sub-voxel offsets and scales in the local neighborhood of the voxel center. This increases local representational capacity by enabling multiple continuous primitives to describe fine structural transitions within one voxel, which is beneficial around tissue boundaries and highly textured regions.

For the coarse branch, given a feature vector $\mathbf{f}_{\boldsymbol{v}_c}^{c} \in \mathbf{F}_{coarse}$ located at the 3D grid coordinate $\boldsymbol{v}_c$, the MLP predicts $m$ ``Structural Gaussian'' $\mathcal{G}_{\boldsymbol{v}_c}^{c} = \{G_{j}^{c}\}_{j=1}^m$. Similarly, for the fine branch, given a feature vector $\mathbf{f}_{\boldsymbol{v}_f}^{f} \in \mathbf{F}_{fine}$ at the upsampled grid coordinate $\boldsymbol{v}_f$, the fine MLP predicts $m$ ``Textural Gaussians" $\mathcal{G}_{\boldsymbol{v}_f}^{f} = \{G_{j}^{f}\}_{j=1}^m$. 
Specifically, for the $j$-th sub-Gaussian in either branch, the projector regresses the 11-dimensional primitive attributes relative to its anchor voxel center $\boldsymbol{v} \in \{\boldsymbol{v}_c, \boldsymbol{v}_f\}$:
\begin{align}
    \boldsymbol{\mu}_{j} &= \boldsymbol{v} + \delta \cdot \tanh(\Delta \boldsymbol{\mu}_{j}), \quad \quad \quad
    \boldsymbol{s}_{j} =  \text{softplus}(\Delta \boldsymbol{s}_{j} - s_{base}), \label{eq:mu_s}\\
    \boldsymbol{q}_{j} &= \frac{\boldsymbol{q}_{raw, j}}{\|\boldsymbol{q}_{raw, j}\|}, \quad \quad \quad \quad \quad \quad \quad
    \alpha_{j} = \text{sigmoid}(\rho_{j} - \alpha_{base}), \label{eq:q_alpha}
\end{align}
where $\delta$ restricts the positional offset to enforce local spatial connectivity, preventing primitive collapse, and $\alpha_j$ denotes the intensity amplitude of the Gaussian.
Crucially, we apply distinct adaptive initialization strategies to enforce the role of each branch. Structural Gaussians are activated with a larger $s_{j}$ and higher $\alpha_{j}$ to cover uniform regions and form the macroscopic anatomical backbone. Textural Gaussians utilize a smaller $s_{j}$ and lower $\alpha_{j}$, specializing in describing microscopic variations. 

To form the complete volumetric representation, we concatenate the output of both branches along the point dimension. Let $N_{coarse} = m \times (HWD)$ and $N_{fine} = m \times (8HWD)$. The final explicit representation is a unified set of $K = N_{coarse} + N_{fine}$ Gaussian primitives:
\begin{equation}
    \mathcal{G} = \bigcup_{\boldsymbol{v}_c} \mathcal{G}_{\boldsymbol{v}_c}^{c} \cup \bigcup_{\boldsymbol{v}_f} \mathcal{G}_{\boldsymbol{v}_f}^{f} = \{G_i\}_{i=1}^{K}.
\end{equation}

\subsection{Differentiable Gaussian Voxelizer}
Unlike standard view synthesis where Gaussians are projected onto a 2D plane via alpha blending along camera rays, our Med3DSR task requires reconstructing a purely 3D scalar field. Since medical volumes represent continuous-valued signals, we formulate the reconstructed continuous intensity function $I(\boldsymbol{x}): \mathbb{R}^3 \to \mathbb{R}$ at any arbitrary spatial query coordinate $\boldsymbol{x}$ as the direct spatial superposition of the Gaussian primitives.

The reconstructed intensity is modeled as the aggregation of contributions from all $K$ Gaussians in the set $\mathcal{G}$. Formally, the voxelizer evaluates the intensity by aggregating contributions from all primitives:
\begin{equation}
I(\boldsymbol{x}) = \sum_{i=1}^{K} \alpha_i \cdot 
\exp\!\left(-\frac{1}{2}(\boldsymbol{x} - \boldsymbol{\mu}_i)^{\top}\boldsymbol{\Sigma}_i^{-1}(\boldsymbol{x} - \boldsymbol{\mu}_i)\right),
\label{eq:reconstruction}
\end{equation}
where $\boldsymbol{\mu}_i$ and $\alpha_i$ denote the center and intensity amplitude of the $i$-th Gaussian, respectively. We use the unnormalized Gaussian kernel as a smooth, distance decaying weighting function. The covariance matrix is parameterized as
$
\boldsymbol{\Sigma}_i = \mathbf{R}(\boldsymbol{q}_i)\,\mathrm{diag}(\boldsymbol{s}_i^2)\,\mathbf{R}(\boldsymbol{q}_i)^\top,
$
with scaling $\boldsymbol{s}_i\in\mathbb{R}_+^3$ and a unit quaternion $\boldsymbol{q}_i$ defining $\mathbf{R}(\boldsymbol{q}_i)$.
For efficiency, we approximate the summation in Eq.~\eqref{eq:reconstruction} by evaluating only primitives within a fixed truncation radius in Mahalanobis space, namely those satisfying $d_M(\boldsymbol{x},\boldsymbol{\mu}_i) < 3$ (a commonly used $3\sigma$ like confidence interval), where
$
d_M(\boldsymbol{x}, \boldsymbol{\mu}_i) = \sqrt{(\boldsymbol{x} - \boldsymbol{\mu}_i)^{\top}\boldsymbol{\Sigma}_i^{-1}(\boldsymbol{x} - \boldsymbol{\mu}_i)}.
$
Primitives outside this neighborhood contribute negligibly and are omitted to reduce computation, while Eq.~\eqref{eq:reconstruction} defines a continuous and differentiable intensity field.

\textbf{Arbitrary-Scale Inference.} The continuous mathematical formulation of Eq.~\eqref{eq:reconstruction} fundamentally decouples the target resolution from the network architecture. During inference, to reconstruct a volume at an arbitrary upsampling scale $\boldsymbol{\kappa}=(\kappa_h,\kappa_w,\kappa_d)$, we simply instantiate a dense spatial coordinate grid of size $(\lceil\kappa_h H\rceil \times \lceil\kappa_w W\rceil \times \lceil\kappa_d D\rceil)$. By querying the continuous intensity function $I(\boldsymbol{x})$ at each corresponding grid coordinate $\boldsymbol{x}$, MedGSSR renders the HR volume at any desired resolution using a single trained model.

\subsection{Optimization Objective}
Since the DGV module enables back-propagation through volumetric rendering, MedGSSR can be trained end-to-end. We supervise the reconstructed volume $\mathbf{V}_{pred}$ (sampled at the target HR grid resolution) against the ground-truth high-resolution volume $\mathbf{V}_{GT}$ using a voxel-wise $L_1$ reconstruction loss:
\begin{equation}
    \mathcal{L}_{rec} = \|\mathbf{V}_{GT} - \mathbf{V}_{pred}\|_1.
\end{equation}
This objective effectively drives the network to recover accurate anatomical structures and textures through the explicit 3D Gaussian representation.

\section{Experiments}

\subsection{Experimental Setup}

\textbf{Datasets.} We conduct comprehensive evaluations on two widely adopted medical imaging modalities: MRI and CT.  For MRI, we use the Medical Segmentation Decathlon (MSD) dataset, which contains high-quality T1-weighted brain volumes for intra-domain training and evaluation. To assess generalization, we select a testing subset from the Human Connectome Project (HCP) dataset, which features different scanner characteristics and population demographics from the MSD dataset. For CT, we train on the MELA dataset, containing diverse chest CT scans, and test generalization on an independent subset of the Ultra-High-Resolution CT (UHRCT) dataset, which captures fine-grained anatomical lung structures. Following standard protocols, low-resolution inputs are generated by downsampling the ground-truth volumes using cubic interpolation.

\textbf{Baselines and Metrics.} MedGSSR is compared against state-of-the-art (SOTA) methods grouped into four paradigms:
(1) Traditional: Trilinear and Cubic interpolation;
(2) Self-supervised: CuNeRF \cite{chen2023cunerf} and 2D-diffusion-guided NAB-GS \cite{noh2025zero};
(3) Neural Implicit: 
INR-based ArSSR \cite{wu2022arbitrary}; and (4) the proposed Feed-Forward 3DGS paradigm (FF-3DGS): MedGSSR.
Reconstruction fidelity is quantified using Peak Signal-to-Noise Ratio (PSNR), Structural Similarity Index (SSIM), and Learned Perceptual Image Patch Similarity (LPIPS).

\textbf{Implementation Details.} We implement MedGSSR in PyTorch. The PAE is built upon a 3-layer 3D U-Net architecture without batch normalization to preserve absolute intensity values. Given an input LR patch (e.g., $1 \times 32 \times 32 \times 32$), the PAE extracts a coarse feature grid of size $128 \times 32^3$. Concurrently, a fine-grained feature grid of size $128 \times 64^3$ is generated via a 3D PixelShuffle upsampling operation. 
The HGP consists of parallel MLPs for the coarse and fine branches. Each MLP contains three fully connected layers with GELU activations, mapping the 128-dimensional features to 11-dimensional Gaussian attributes
($\alpha, \boldsymbol{s}, \boldsymbol{\mu}, \boldsymbol{q}$) 
for $m=4$ sub-voxels. To explicitly enforce the structure-texture decoupling, we apply distinct initializations for the two branches: Structural Gaussians use a smaller $s_{base}$ 4.0 and amplitude bias $\alpha_{base}$ 2.0 to model dense macroscopic structures, whereas Textural Gaussians are initialized with a larger $s_{base}$ 6.0 and amplitude bias $\alpha_{base}$ 4.0 to capture sparse high-frequency edges. The entire framework is trained end-to-end for 1,000,000 iterations using the Adam optimizer. The learning rate is initialized at $1 \times 10^{-4}$ and decays logarithmically to $1 \times 10^{-5}$. All experiments are conducted on four NVIDIA RTX 4090 GPUs.

\subsection{Main Results}

\begin{table}[t] 
\centering
\caption{Comparison of intra-domain 3D super-resolution performance on the MSD (MRI) dataset. Models are both trained and evaluated on the MSD dataset.}
\label{tab:main_mri}
\resizebox{\textwidth}{!}{
\begin{tabular}{llccccccccc}
\toprule
\multirow{2}{*}{Paradigm} & \multirow{2}{*}{Methods} & \multicolumn{3}{c}{$2\times$} & \multicolumn{3}{c}{$3\times$} & \multicolumn{3}{c}{$4\times$} \\
\cmidrule(lr){3-5} \cmidrule(lr){6-8} \cmidrule(lr){9-11}
& & PSNR $\uparrow$ & SSIM $\uparrow$ & LPIPS $\downarrow$ & PSNR $\uparrow$ & SSIM $\uparrow$ & LPIPS $\downarrow$ & PSNR $\uparrow$ & SSIM $\uparrow$ & LPIPS $\downarrow$ \\
\midrule
\multirow{2}{*}{Traditional} & Trilinear & 33.31 & 0.9670 & 0.1134 & 30.77 & 0.9377 & 0.2026 & 29.08 & 0.9100 & 0.2558 \\
& Cubic & 33.99 & 0.9725 & 0.0981 & 31.13 & 0.9422 & 0.2083 & 29.27 & 0.9118 & 0.2711 \\
\midrule
Self-supervised & CuNeRF \cite{chen2023cunerf} & 32.60 & 0.9715 & 0.0513 & 30.07 & 0.9524 & 0.0878 & 28.26 & 0.9309 & 0.1271 \\
\midrule
Neural Implicit & ArSSR \cite{wu2022arbitrary} & 32.87 & 0.9742 & 0.0581 & 29.56 & 0.9442 & 0.0730 & 28.51 & 0.9315 & \textbf{0.0853} \\ \midrule
FF-3DGS & MedGSSR & \textbf{35.91} & \textbf{0.9821} & \textbf{0.0337} & \textbf{33.97} & \textbf{0.9700} & \textbf{0.0706} & \textbf{32.10} & \textbf{0.9541} & 0.1246 \\
\bottomrule
\end{tabular}
}
\end{table}

\begin{table}[t]
\centering
\caption{Comparison of intra-domain 3D super-resolution performance on the MELA (CT) dataset. Models are both trained and evaluated on the MELA dataset.}
\label{tab:main_ct}
\resizebox{\textwidth}{!}{
\begin{tabular}{llccccccccc}
\toprule
\multirow{2}{*}{Paradigms} & \multirow{2}{*}{Methods} & \multicolumn{3}{c}{$2\times$} & \multicolumn{3}{c}{$3\times$} & \multicolumn{3}{c}{$4\times$} \\
\cmidrule(lr){3-5} \cmidrule(lr){6-8} \cmidrule(lr){9-11}
& & PSNR $\uparrow$ & SSIM $\uparrow$ & LPIPS $\downarrow$ & PSNR $\uparrow$ & SSIM $\uparrow$ & LPIPS $\downarrow$ & PSNR $\uparrow$ & SSIM $\uparrow$ & LPIPS $\downarrow$ \\
\midrule
\multirow{2}{*}{Traditional} & Trilinear & 37.51 & 0.9530 & 0.1561 & 34.83 & 0.9166 & 0.2302 & 33.84 & 0.8843 & 0.3291 \\
& Cubic & 37.64 & 0.9636 & 0.1179 & 35.14 & 0.9237 & 0.2325 & 34.09 & 0.8888 & 0.3263 \\
\midrule
\multirow{2}{*}{Self-supervised} & CuNeRF \cite{chen2023cunerf} & 37.12 & 0.9654 & 0.1201 & 35.42 & 0.9235 & 0.1863 & 33.94 & 0.8957 & 0.2456 \\
& NAB-GS \cite{noh2025zero} & - & - & - & - & - & - & 34.13 & \textbf{0.9518} & - \\
\midrule
Neural Implicit & ArSSR \cite{wu2022arbitrary} & 38.53 & 0.9644 & 0.0914 & 36.02 & 0.9413 & 0.1576 & 34.63 & 0.9244 & 0.2261 \\ \midrule
FF-3DGS & MedGSSR & \textbf{42.08} & \textbf{0.9738} & \textbf{0.0705} & \textbf{39.95} & \textbf{0.9558} & \textbf{0.1193} & \textbf{37.31} & 0.9362 & \textbf{0.1472} \\
\bottomrule
\end{tabular}
}
\end{table}
\begin{figure}[t]
    \centering
    \includegraphics[width=\textwidth]{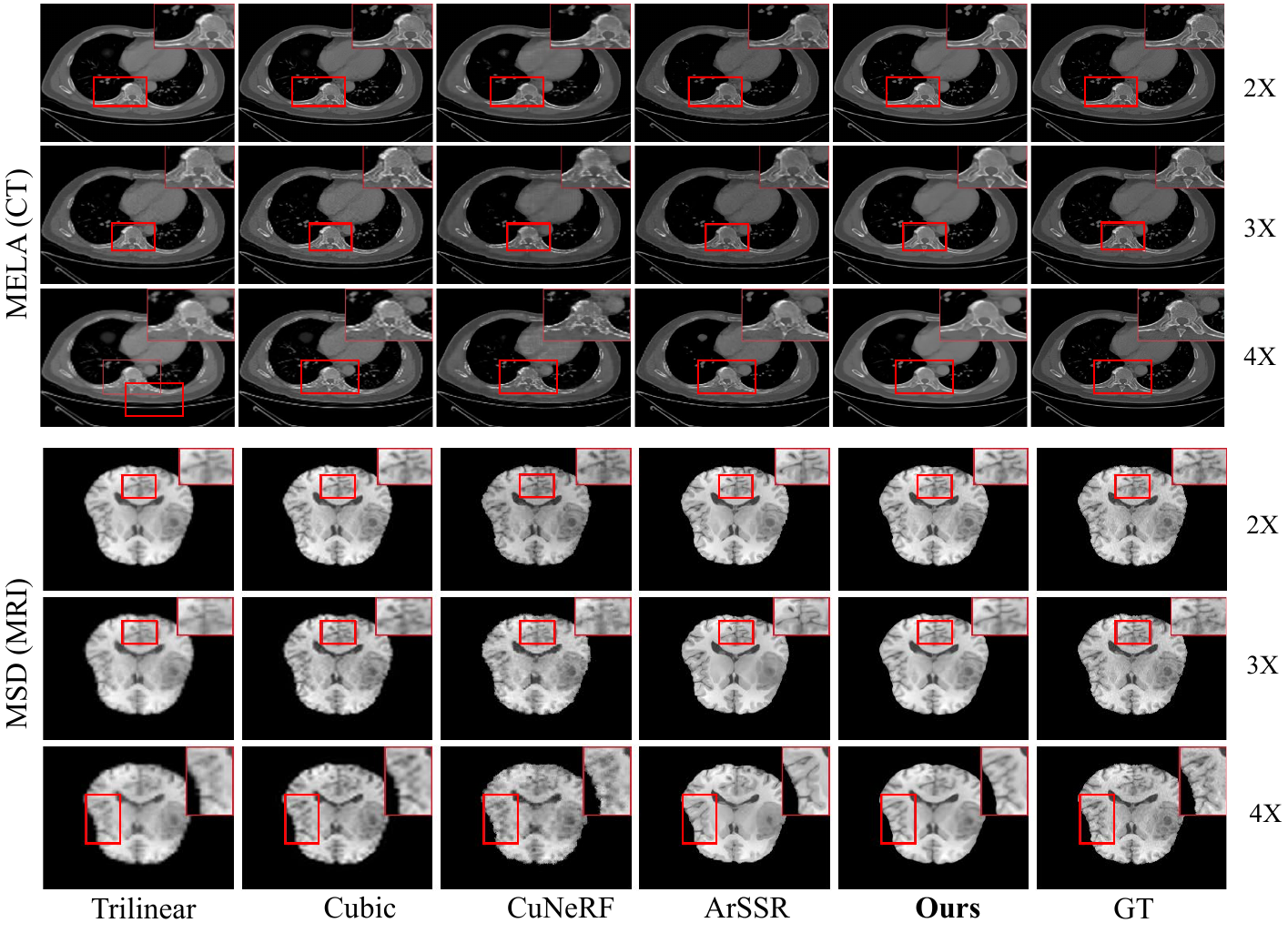}
    \vspace{-0.6cm}
    \caption{\textbf{Qualitative comparison of intra-domain 3DSR on MELA (CT) and MSD (MRI).} Axial views are shown at $2\times$, $3\times$ and $4\times$ scales. MedGSSR better recovers fine-grained skeletal and vascular structures, avoiding the smoothing artifacts of ArSSR and the noise commonly observed in other methods.}
    \label{fig:visual_ct}
\end{figure}

\begin{table}[t]
\centering
\caption{Comparison of cross-domain 3D super-resolution performance on the HCP (MRI). Models are trained on the MSD dataset and evaluated on the HCP dataset.}
\label{tab:cross_mri}
\resizebox{\textwidth}{!}{
\begin{tabular}{llccccccccc}
\toprule
\multirow{2}{*}{Paradigms} & \multirow{2}{*}{Methods} & \multicolumn{3}{c}{$2\times$} & \multicolumn{3}{c}{$3\times$} & \multicolumn{3}{c}{$4\times$} \\
\cmidrule(lr){3-5} \cmidrule(lr){6-8} \cmidrule(lr){9-11}
& & PSNR $\uparrow$ & SSIM $\uparrow$ & LPIPS $\downarrow$ & PSNR $\uparrow$ & SSIM $\uparrow$ & LPIPS $\downarrow$ & PSNR $\uparrow$ & SSIM $\uparrow$ & LPIPS $\downarrow$ \\
\midrule
\multirow{2}{*}{Traditional} & Trilinear & 36.45 & 0.9425 & 0.1048 & 33.75 & 0.9221 & 0.1634 & 30.76 & 0.9026 & 0.2675 \\
& Cubic & 36.47 & 0.9588 & 0.0885 & 34.01 & 0.9289 & 0.1454 & 31.86 & 0.9125 & 0.2483 \\
\midrule
Self-supervised & CuNeRF \cite{chen2023cunerf} & 36.91 & 0.9716 & 0.0724 & 34.48 & 0.9547 & 0.1362 & 33.12 & 0.9382 & 0.1587 \\
\midrule
Neural Implicit  & ArSSR \cite{wu2022arbitrary} & 36.07 & 0.9655 & \textbf{0.0615} & 33.56 & 0.9429 & \textbf{0.0892} & 31.87 & 0.9220 & \textbf{0.1141} \\ \midrule
FF-3DGS & MedGSSR & \textbf{40.81} & \textbf{0.9849} & 0.0674 & \textbf{37.94} & \textbf{0.9702} & 0.0945 & \textbf{35.84} & \textbf{0.9533} & 0.1253 \\
\bottomrule
\end{tabular}
}
\end{table}

\begin{table}[t]
\centering
\caption{Comparison of cross-domain 3D super-resolution performance on the UHRCT (CT). Models are trained on the MELA dataset and evaluated on the UHRCT dataset.}
\label{tab:cross_ct}
\resizebox{\textwidth}{!}{
\begin{tabular}{llccccccccc}
\toprule
\multirow{2}{*}{Paradigms} & \multirow{2}{*}{Methods} & \multicolumn{3}{c}{$4\times$} & \multicolumn{3}{c}{$6\times$} & \multicolumn{3}{c}{$8\times$} \\
\cmidrule(lr){3-5} \cmidrule(lr){6-8} \cmidrule(lr){9-11}
& & PSNR $\uparrow$ & SSIM $\uparrow$ & LPIPS $\downarrow$ & PSNR $\uparrow$ & SSIM $\uparrow$ & LPIPS $\downarrow$ & PSNR $\uparrow$ & SSIM $\uparrow$ & LPIPS $\downarrow$ \\
\midrule
\multirow{2}{*}{Traditional} & Trilinear & 31.96 & 0.9404 & 0.1051 & 26.73 & 0.9027 & 0.1706 & 23.96 & 0.8436 & 0.2605 \\
& Cubic & 32.88 & 0.9407 & 0.0898 & 28.51 & 0.9189 & 0.1654 & 26.07 & 0.8579 & 0.2517 \\
\midrule
Self-supervised & CuNeRF \cite{chen2023cunerf} & 38.03 & 0.9628 & 0.1252 & 33.76 & 0.9164 & 0.2465 & 30.50 & 0.8690 & 0.3510 \\
\midrule
Neural Implicit & ArSSR \cite{wu2022arbitrary} & 35.92 & 0.9471 & 0.0918 & 32.88 & 0.9332 & 0.1191 & 29.38 & 0.8994 & 0.2145 \\ \midrule
FF-3DGS & MedGSSR & \textbf{41.20} & \textbf{0.9818} & \textbf{0.0468} & \textbf{37.74} & \textbf{0.9698} & \textbf{0.0765} & \textbf{35.24} & \textbf{0.9233} & \textbf{0.1253} \\
\bottomrule
\end{tabular}
}
\end{table}

\begin{figure}[t]
    \centering
    \includegraphics[width=\textwidth]{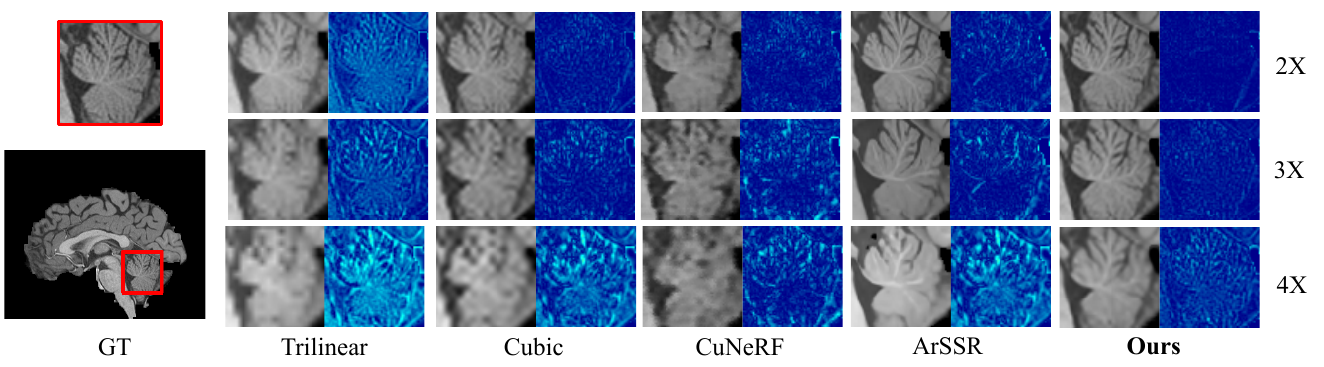}
    \vspace{-0.6cm}
    \caption{\textbf{Qualitative comparison of cross-domain 3DSR on the HCP (MRI).} All models are trained on the MSD and evaluated on the unseen HCP. The error maps (dark blue indicates lower error) highlight that MedGSSR produces the sharpest anatomical details with minimal reconstruction deviation compared to the baselines.}    
    \label{fig:visual_hcp}
\end{figure}

\textbf{Quantitative Analysis.}
We first assess the reconstruction fidelity of MedGSSR within the primary MRI and CT domains. As reported in Table \ref{tab:main_mri}, MedGSSR establishes a new state-of-the-art across spatial scales on the MSD dataset. At the highly challenging $4\times$ upsampling scale, our method achieves a PSNR of 32.10 dB, delivering a remarkable improvement (+3.59 dB) over the neural implicit method ArSSR. These results quantitatively prove that representing anatomy with explicit, localized Gaussian primitives successfully bypasses the spectral bias of coordinate-based MLPs, enabling the network to reconstruct high-frequency structural variations without the typical over-smoothing penalty.
Table \ref{tab:main_ct} details the corresponding results for the CT modality on the MELA dataset. Consequently, at the challenging $4\times$ scale, our method achieves the highest geometric accuracy (37.31 dB PSNR), outperforming ArSSR by +2.68 dB. Additionally, for the diffusion-guided method NAB-GS, due to the unavailability of its open-source code, we directly report the $4\times$ performance quoted from their original paper. Notably, MedGSSR still surpasses this reported performance by a substantial +3.18 dB margin. These results provide strong evidence that learning a 3D-native feed-forward mapping directly from voxel grids is fundamentally superior to paradigms constrained by isolated optimization, coordinate-based mappings, or 2D proxy supervision.

\textbf{Qualitative Visualizations.}
Fig. \ref{fig:visual_ct} shows qualitative comparisons on MELA (CT) and MSD (MRI) in the axial plane at $2\times$, $3\times$, and $4\times$ upsampling factors. The self-supervised CuNeRF and the neural implicit ArSSR produce noticeable blurring, and ArSSR yields over-smoothed tissue boundaries. In contrast, MedGSSR reconstructs sharper anatomical structures, such as clearer skeletal edges and finer vessels, supporting the effectiveness of our dual-resolution structure-texture decoupling. Additional visualizations are provided in the \textbf{Supplementary Material}.

\subsection{Generalization Results}
A fundamental limitation of isolated optimization methods is their inability to generalize to unseen datasets. To assess the cross-dataset generalizability of MedGSSR, we evaluate models trained on the source datasets (MSD/MELA) directly on held-out target datasets (HCP for MRI and UHRCT for CT), without any fine-tuning.
Tables \ref{tab:cross_mri} and \ref{tab:cross_ct} summarize these generalization results. As observed, the method ArSSR experiences notable performance degradation when encountering unseen datasets. Conversely, MedGSSR maintains exceptional reconstruction fidelity, securing a PSNR improvement of 3.97 dB over ArSSR at the $4\times$ scale on the HCP dataset. 
More impressively, on the UHRCT dataset (Table \ref{tab:cross_ct}), MedGSSR is evaluated at extreme upsampling scales ($4\times, 6\times, 8\times$). Even at an $8\times$ resolution gap, our method yields a PSNR surge (+5.86 dB) over ArSSR and a gain (+4.74 dB) over CuNeRF. 
As visualized in Fig.~\ref{fig:visual_hcp}, baseline methods tend to produce blurred or noisy reconstructions under domain shift. In contrast, MedGSSR better preserves structurally consistent cortical folds and tissue boundaries. These results support that our model learns transferable priors from the source data, enabling robust feed-forward recovery of high-frequency details on unseen domains. Additional visualizations are provided in the \textbf{Supplementary Material}.

\begin{figure*}[t]
    \centering
    \includegraphics[width=\textwidth]{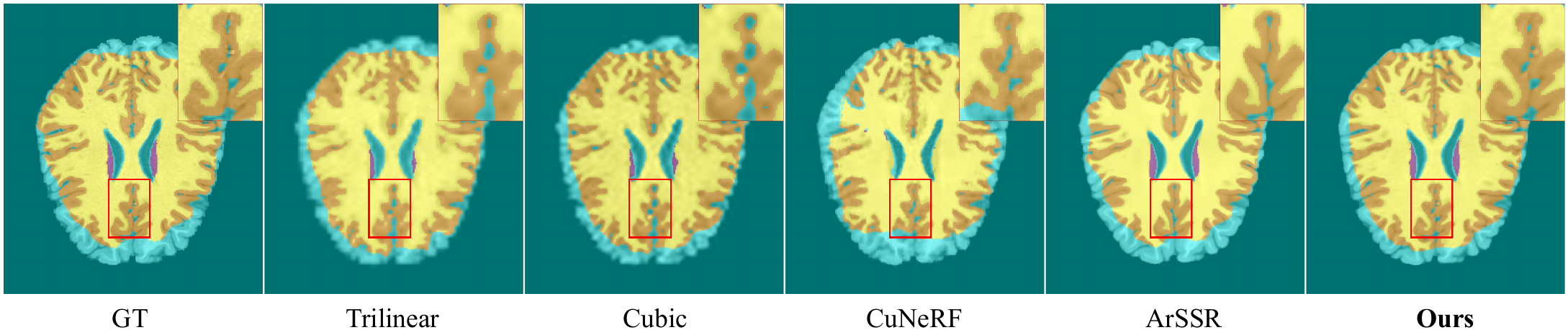}
    \vspace{-0.6cm}
    \caption{\textbf{Downstream brain tissue segmentation on the HCP dataset ($4\times$ SR).} The structural fidelity provided by MedGSSR translates to precise automated boundary delineation, avoiding the tissue-bleeding artifacts observed in the others.}
    \label{fig:visual_seg}
\end{figure*}

\textbf{Clinical Validation.} To verify that the visual improvements of MedGSSR translate into actual clinical utility, we perform an automated brain tissue segmentation task using a pre-trained nnUNet model. The segmentation is conducted on the $4\times$ upsampled MRI volumes from the HCP dataset. 
Table \ref{tab:dice_scores} reports the Dice Similarity Coefficients across five brain regions (CSF, GM, WM, dGM, BS+CB). MedGSSR achieves the highest average Dice score (0.8178), outperforming CuNeRF (which completely collapses to 0.3618 due to severe artifacts) and ArSSR (+3.4\% relative increase). As visualized in Fig. \ref{fig:visual_seg}, the volumes reconstructed by ArSSR exhibit blurry boundaries, leading to under-segmentation and bleeding at tissue margins. Thanks to our hierarchical structure-texture decoupling, MedGSSR restores sharp anatomical edges, yielding segmentation masks that align with the original ground truth. Additional visualizations are provided in the \textbf{Supplementary Material}.
\begin{table}[t]
\centering
\caption{Comparison of Dice scores. The segmentation is performed using a pre-trained nnUNet on $4\times$ super-resolved MRI volumes (HCP). Best results are highlighted in bold.}
\label{tab:dice_scores}
{
\renewcommand{\arraystretch}{0.85}
\begin{tabular*}{\textwidth}{@{\extracolsep{\fill}}lcccccc}
\toprule
Methods & CSF & GM & WM & dGM & BS+CB & Average \\
\midrule
Trilinear & 0.4937 & 0.7265 & 0.8617 & 0.8098 & 0.6701 & 0.7124 \\
Cubic     & 0.5410 & 0.7616 & 0.8860 & 0.8616 & \textbf{0.8518} & 0.7804 \\
CuNeRF \cite{chen2023cunerf}   & 0.3668 & 0.5330 & 0.7347 & 0.1285 & 0.0458 & 0.3618 \\
ArSSR \cite{wu2022arbitrary}    & \textbf{0.6782} & 0.8318 & 0.9187 & 0.9140 & 0.5612 & 0.7908 \\
\midrule
Ours (MedGSSR)      & 0.5779 & \textbf{0.8406} & \textbf{0.9295} & \textbf{0.9407} & 0.8006 & \textbf{0.8178} \\
\bottomrule
\end{tabular*}
}
\end{table}

\subsection{Ablation Study}
Unless otherwise specified, all ablation experiments are conducted under the intra-domain setting on the HCP MRI dataset for $4\times$ 3D SR.

\begin{figure*}[h]
    \centering
    \includegraphics[width=0.75\textwidth]{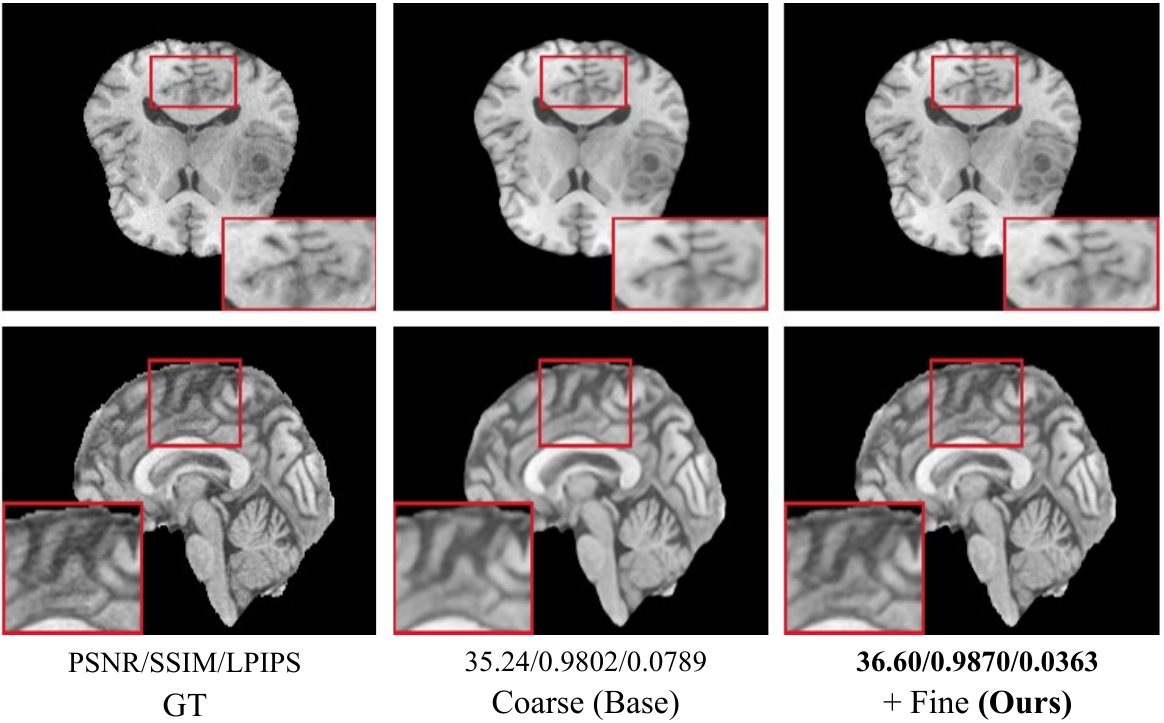}
    \caption{\textbf{Ablation on Structure-Texture Decoupling on the HCP dataset ($4\times$ SR).} The coarse branch captures global anatomy, and adding the fine branch restores high-frequency textures, improving sharpness and quantitative metrics.}
    \label{fig:ablation_visual}
\end{figure*}

\textbf{Structure-Texture Decoupling.} To validate the design of MedGSSR, we study the effect of dual-resolution structure-texture decoupling. Fig.~\ref{fig:ablation_visual} reports a qualitative and quantitative comparison between a coarse-only variant and the full model. The coarse branch captures global anatomy (35.24~dB PSNR, 0.9802 SSIM), but fails to recover fine-scale boundaries, resulting in blurred cortical folds. Adding the fine branch explicitly models high-frequency residuals, improving reconstruction quality to 36.60~dB PSNR and 0.9870 SSIM, while reducing LPIPS from 0.0789 to 0.0363. These results indicate that the decoupled hierarchical prediction is important for preserving fine anatomical details.

\textbf{Impact of Sub-voxel Count.} 
The hyperparameter $m$ determines the number of continuous Gaussian spheres split from each discrete latent feature location. 
We examine its effect on reconstruction quality in Table \ref{tab:ablation_m}. 
With $m=1$, the limited geometric capacity is insufficient to model complex anatomical boundaries, resulting in inferior performance (35.98~dB PSNR). Increasing $m$ to 2 and 4 improves both fidelity and perceptual quality, reaching the best results at $m=4$ (36.60~dB PSNR). Further increasing to $m=8$ brings marginal saturation and slightly worse metrics (36.52~dB PSNR). We therefore set $m=4$ by default to balance reconstruction quality and computational cost.

\begin{table}[t]
\centering
\caption{Ablation on sub-voxel count $m$ under intra-domain setting on the HCP dataset (4$\times$ SR). Setting $m=4$ delivers the best balance between accuracy and computational cost. Best results are highlighted in bold.}
\label{tab:ablation_m}
{
\renewcommand{\arraystretch}{0.85}
\begin{tabular*}{\linewidth}{@{\extracolsep{\fill}}lccc}
\toprule
Settings & PSNR $\uparrow$ & SSIM $\uparrow$ & LPIPS $\downarrow$ \\
\midrule
$m = 1$ & 35.98 & 0.9760 & 0.0423 \\
$m = 2$ & 36.06 & 0.9771 & 0.0388 \\
$m = 4$ (Default) & \textbf{36.60} & \textbf{0.9870} & \textbf{0.0363} \\
$m = 8$ & 36.52 & 0.9862 & 0.0365 \\
\bottomrule
\end{tabular*}
}
\end{table}

\textbf{Effect of Gaussian Support, Frequency Supervision, and Data Scale.}
We further investigate the effects of the Gaussian support radius, an auxiliary frequency-domain objective, and training data scale on the HCP dataset under the cross-domain setting for $4\times$ 3D SR. As shown in Table~\ref{tab:ablation_mri}, adding the FFT loss improves LPIPS from 0.1253 to 0.1078, suggesting better perceptual high-frequency recovery, but slightly decreases PSNR and SSIM. This indicates a trade-off between perceptual sharpness and voxel-wise fidelity. In contrast, reducing the truncation radius to $1\sigma$ causes a large degradation across all metrics, confirming that sufficient Gaussian support is necessary for aggregating neighboring primitives and maintaining a continuous volumetric field. We also examine the robustness of MedGSSR under reduced training data. The results show that MedGSSR remains robust when using 75\% of the training data. MedGSSR remains competitive with 75\% training data, achieving 35.49 dB PSNR and 0.9512 SSIM, while performance drops more noticeably with 50\% data. Additional ablations are provided in the \textbf{Supplementary Material}.

\begin{table}[t]
\centering
\caption{Ablation study on effects of frequency-domain supervision, Gaussian truncation radius, and training data under the cross-domain setting on the HCP dataset ($4\times$ SR). Best results are highlighted in bold.}
\label{tab:ablation_mri} 
{ \renewcommand{\arraystretch}{0.85} 
\begin{tabular*}{\linewidth}{@{\extracolsep{\fill}}lccc} 
\toprule Variants & PSNR $\uparrow$ & SSIM $\uparrow$ & LPIPS $\downarrow$ \\ \midrule 
Ours & $\mathbf{35.84}$ & $\mathbf{0.9533}$ & ${0.1253}$ \\
Ours $+$ FFT Loss & ${35.72}$ & $0.9472$ & $\mathbf{0.1078}$ \\
Ours w/ $1\sigma$ truncation & $31.05$ & $0.9142$ & $0.1608$ \\ 
Ours w/ 50\% training data & $34.26$ & $0.9405$ & $0.1452$ \\ 
Ours w/ 75\% training data & $35.49$ & ${0.9512}$ & $0.1283$ \\ 
\bottomrule 
\end{tabular*} } 
\vspace{-0.3cm}
\end{table}

\section{Conclusion}
In this paper, we present MedGSSR, a feed-forward framework that formulates medical 3D super-resolution as a generalized volumetric reconstruction problem. By leveraging explicit 3D Gaussian Splatting as continuous basis functions, MedGSSR removes the need for isolated per-volume optimization in self-supervised pipelines and alleviates over-smoothing effects commonly observed in implicit neural representations. To address anatomical heterogeneity, we introduce a Hierarchical Gaussian Projector with sub-voxel decomposition, which decouples global structural preservation from fine-scale detail refinement. Moreover, the proposed Differentiable Gaussian Voxelizer enables end-to-end supervision directly in the 3D domain, avoiding reliance on 2D proxy projections and reducing geometric ambiguity. Extensive experiments across MRI and CT benchmarks demonstrate strong medical super-resolution with high fidelity, and robust cross-dataset generalization under domain shift. Finally, the improved structural accuracy of MedGSSR consistently benefits downstream clinical tasks, highlighting its potential as an efficient and reliable foundation for volumetric medical image enhancement.

%
%
\bibliographystyle{splncs04}
\bibliography{arxiv}

\clearpage

\setcounter{section}{0}
\renewcommand{\thesection}{\Alph{section}}
\makeatletter
\@ifundefined{theHsection}{}{\renewcommand{\theHsection}{supp.\Alph{section}}}
\makeatother

\section{Additional Visual Comparisons for Intra-Domain 3D Super-Resolution}
To complement the quantitative results in the main paper, we provide additional visual comparisons for intra-domain 3D super-resolution on both MRI and CT data. For each example, we show three anatomical planes, namely axial, coronal, and sagittal views, together with enlarged local regions to better highlight subtle structural differences. Results are presented at 2$\times$, 3$\times$, and 4$\times$ super-resolution scales.

\begin{figure*}[htbp]
\centering
\includegraphics[width=\textwidth]{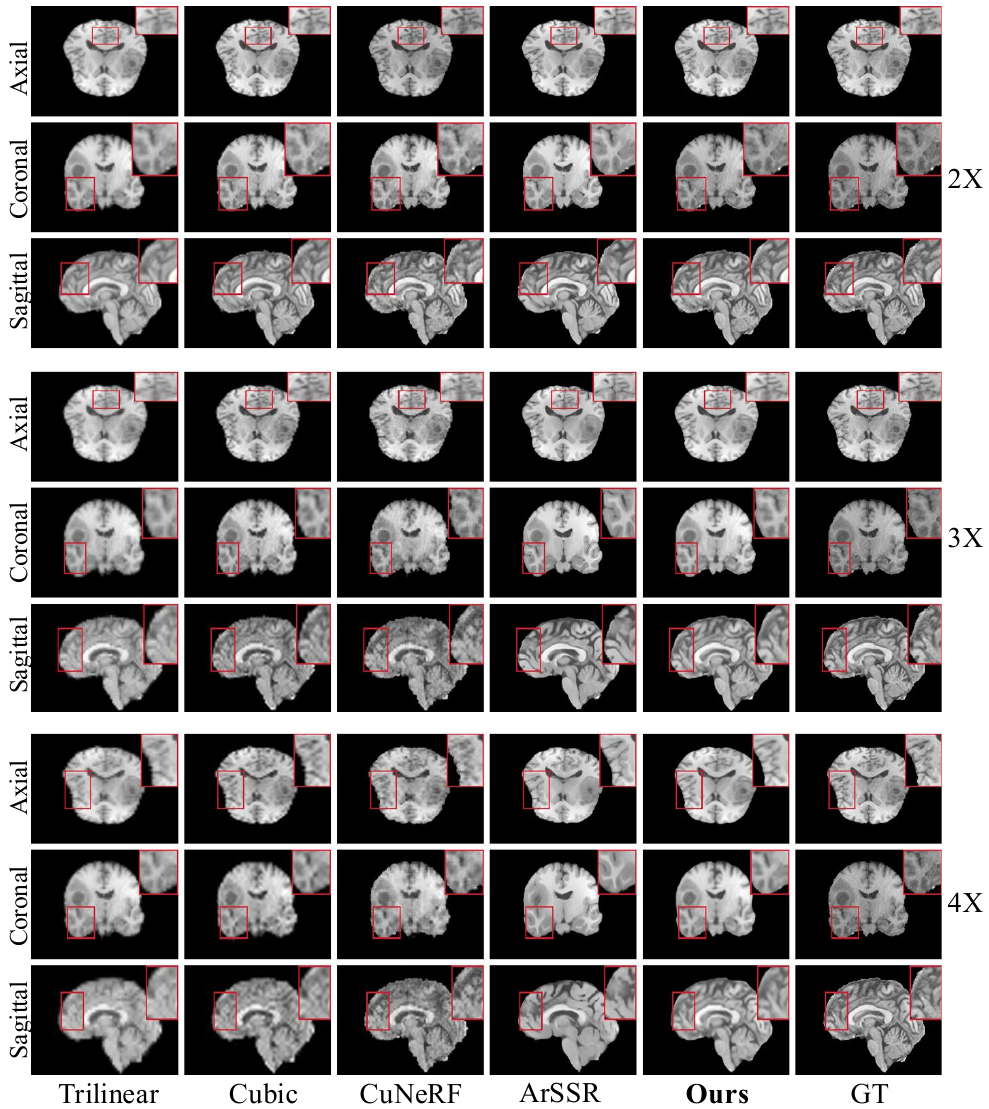}
\caption{Additional visual comparisons for intra-domain 3D super-resolution on the MSD (MRI) dataset. Results are shown across axial, coronal, and sagittal views at 2$\times$, 3$\times$, and 4$\times$ upsampling scales, with zoomed-in regions for detailed inspection. From left to right: Trilinear, Cubic, CuNeRF, ArSSR, Ours, and GT. MedGSSR better preserves cortical folds, tissue boundaries, and local anatomical details than competing methods.}
\label{fig:supp_msd}
\end{figure*}

Figure~\ref{fig:supp_msd} shows the intra-dataset MRI super-resolution results on the Medical Segmentation Decathlon (MSD) dataset~\cite{antonelli2022medical}. As the upsampling factor increases, the interpolation baselines become increasingly blurry and fail to preserve clear anatomical boundaries. Although CuNeRF and ArSSR recover part of the global contrast, they still exhibit noticeable over-smoothing and lose fine cortical folding patterns, especially at 4$\times$. In contrast, MedGSSR restores sharper structures and more faithful local details across all three anatomical planes, with clearer recovery of thin folds and tissue boundaries in the zoomed regions. These visual results are consistent with the quantitative gains reported in the main paper and further demonstrate the advantage of Gaussian-based continuous volumetric modeling for MRI super-resolution.

\begin{figure*}[htbp]
\centering
\includegraphics[width=\textwidth]{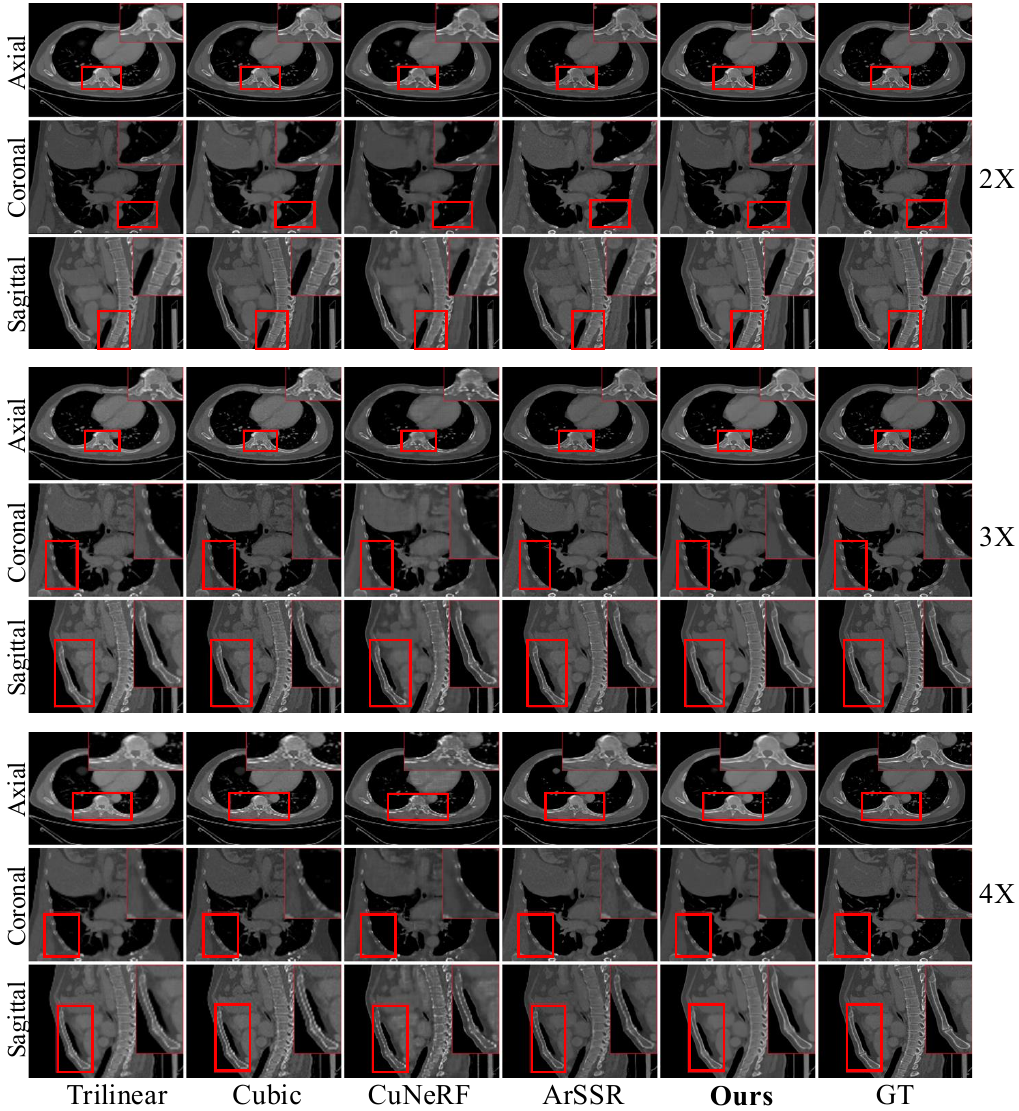}
\caption{Additional visual comparisons for intra-domain 3D super-resolution on the MELA (CT) dataset. Results are shown across axial, coronal, and sagittal views at 2$\times$, 3$\times$, and 4$\times$ upsampling scales, with zoomed-in regions for detailed inspection. From left to right: Trilinear, Cubic, CuNeRF, ArSSR, Ours, and GT. Compared with the baselines, MedGSSR produces clearer boundaries and more faithful recovery of fine anatomical structures across all super-resolution.}
\label{fig:supp_mela}
\end{figure*}

Figure~\ref{fig:supp_mela} presents the corresponding intra-dataset CT super-resolution results on the MELA dataset \cite{mela}. Compared with MRI, CT images contain both strong macroscopic structures and subtle high-frequency details, which makes accurate super-resolution particularly challenging. As shown in the figure, Trilinear and Cubic interpolation produce blurred boundaries and degraded local contrast, while CuNeRF and ArSSR still struggle to faithfully recover fine structures in challenging regions. By comparison, MedGSSR better preserves anatomical boundaries and restores clearer local details across different scales and planes, while introducing fewer artifacts. The improvements are especially visible in the magnified regions, where our method produces reconstructions that are consistently closer to the ground truth.

Overall, these visual comparisons further support the main paper by showing that the proposed Gaussian-based super-resolution framework consistently preserves both global anatomical structure and local high-frequency detail under intra-dataset settings.

\section{Additional Visual Comparisons for Cross-Domain 3D Super-Resolution Generalization}
To complement the cross-domain quantitative results reported in the main paper, we provide additional visual comparisons on unseen MRI and CT datasets. These visualizations are intended to further illustrate the robustness of the proposed method under distribution shift.

\begin{figure*}[htbp]
\centering
\includegraphics[width=\textwidth]{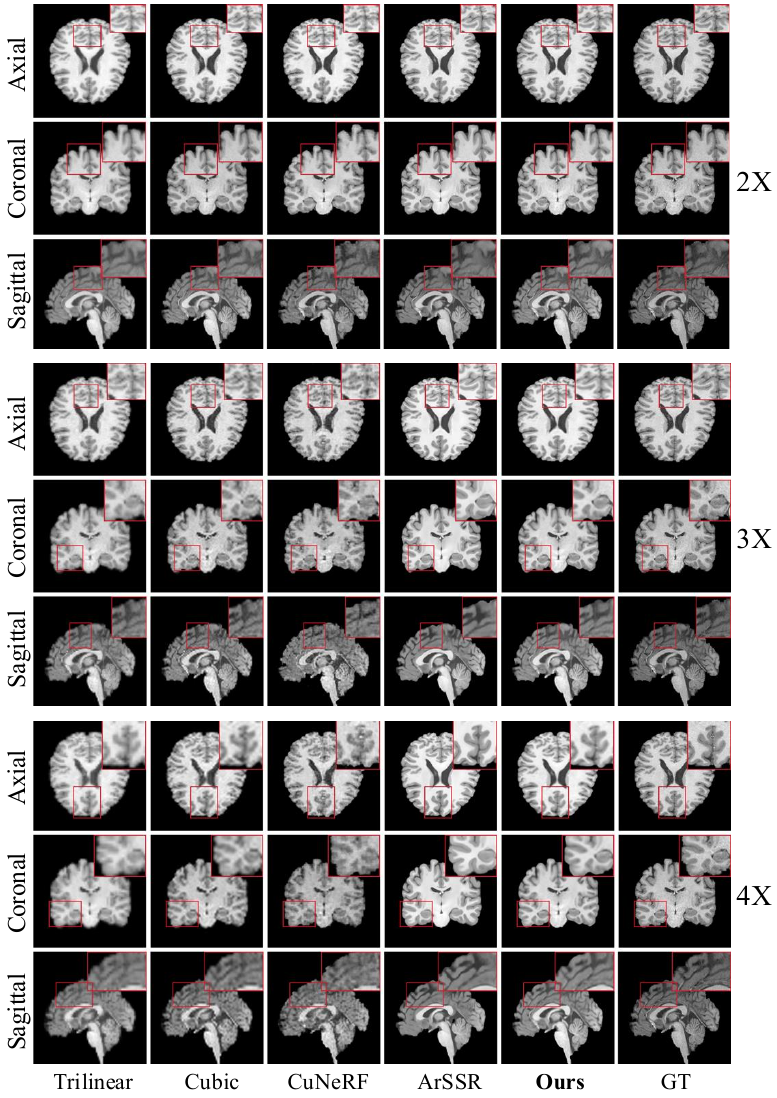}
\caption{Additional visual comparisons for cross-domain 3D super-resolution on the unseen HCP (MRI) dataset~\cite{van2013wu}. Models are trained exclusively on the MSD dataset and evaluated at 2$\times$, 3$\times$, and 4$\times$ upsampling scales. Results are shown across axial, coronal, and sagittal views, with zoomed-in regions for detailed inspection. Compared with competing methods, MedGSSR better preserves tissue boundaries, cortical structures, and local anatomical details under dataset shift.}
\label{fig:supp_hcp}
\end{figure*}

Figure~\ref{fig:supp_hcp} presents additional cross-dataset MRI super-resolution results on the unseen Human Connectome Project (HCP) dataset~\cite{van2013wu}. All models are trained exclusively on the MSD dataset and directly evaluated on HCP at 2$\times$, 3$\times$, and 4$\times$ upsampling scales. As shown across the axial, coronal, and sagittal views, interpolation based baselines become increasingly blurry as the scale factor increases, while CuNeRF and ArSSR exhibit noticeable degradation in fine anatomical structures under dataset shift. In contrast, MedGSSR preserves clearer tissue boundaries, more coherent cortical patterns, and more faithful local details across all three planes. These visual observations are consistent with the cross-dataset quantitative results in the main paper and further support the strong generalization capability of the proposed Gaussian-based super-resolution framework.

\begin{figure*}[htbp]
\centering
\includegraphics[width=\textwidth]{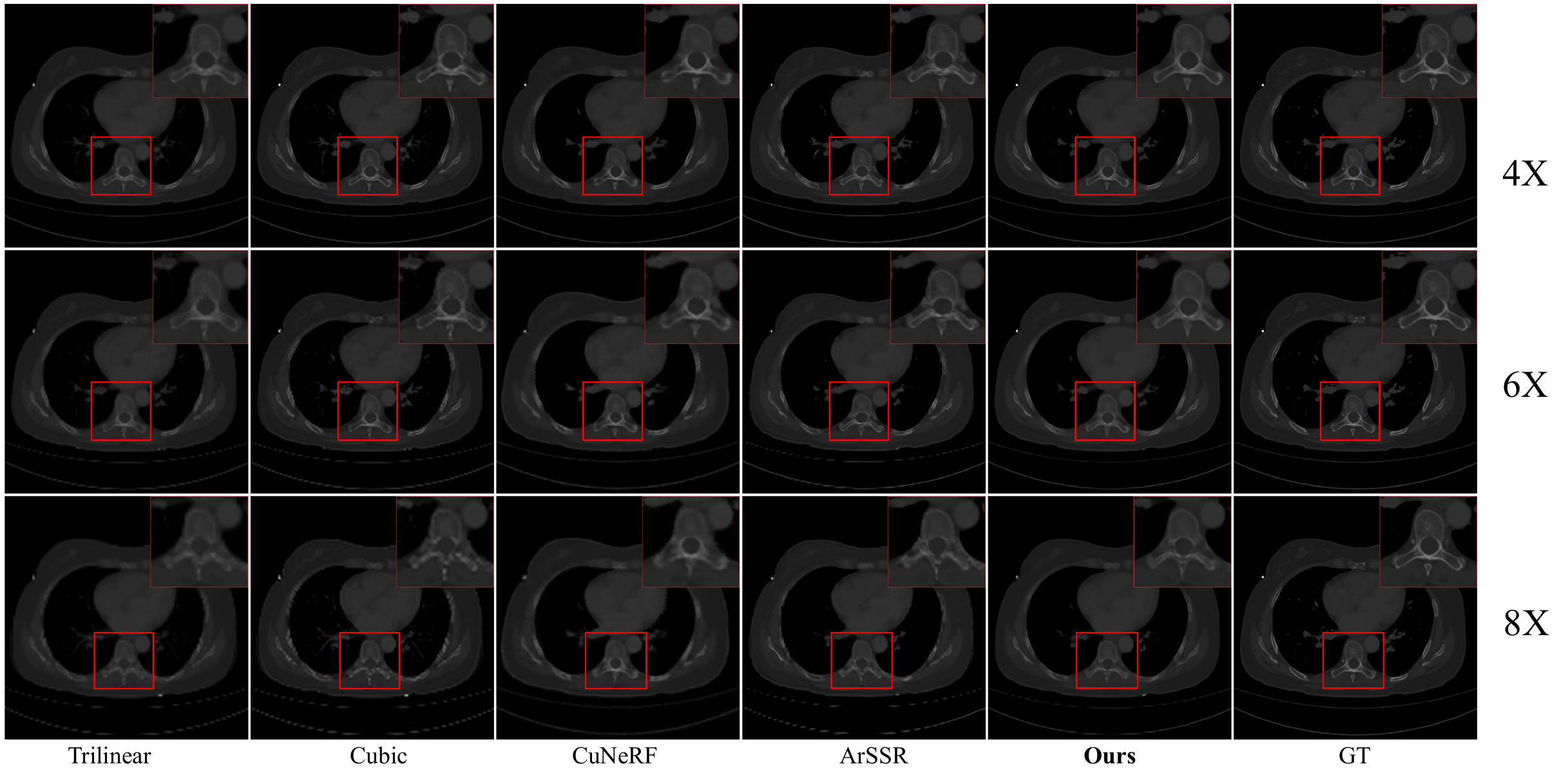}
\caption{Additional visual comparisons for cross-dataset 3D super-resolution on the unseen UHRCT (CT) dataset~\cite{chu2023topology}. Models are trained on MELA and evaluated at 4$\times$, 6$\times$, and 8$\times$ upsampling scales. Following the benchmark protocol, results are shown on the axial plane, with zoomed-in regions for detailed inspection. MedGSSR remains more robust than competing methods as the resolution gap increases, preserving clearer anatomical boundaries and finer local structures.}
\label{fig:supp_uhrct}
\end{figure*}

Figure~\ref{fig:supp_uhrct} shows the corresponding cross-dataset CT results on the unseen Ultra-High-Resolution CT (UHRCT) dataset~\cite{chu2023topology}, using models trained only on the MELA dataset. Following the evaluation protocol of this benchmark, we visualize the axial plane, which is also the most reliable plane for qualitative comparison under the native acquisition setting. In addition to the 4$\times$ case considered in the main paper, we further include more challenging 6$\times$ and 8$\times$ visual comparisons as an additional stress test of arbitrary-scale generalization. As the resolution gap increases, the competing methods show progressively stronger blurring and structural degradation, whereas MedGSSR still preserves clearer anatomical boundaries and more faithful local structures. These results further demonstrate that the proposed method remains robust under both dataset shift and large upsampling factors.

\section{Extended Baseline and Efficiency Analysis on Cross-Domain HCP MRI}
To further validate the cross-dataset generalization and computational efficiency of MedGSSR, we provide an extended comparison on the unseen HCP MRI dataset under the $4\times$ setting. In addition to the baselines reported in the main paper, we include two transformer-based 3D super-resolution methods, MTVNet \cite{hoeg2024mtvnet} and SuperFormer \cite{forigua2022superformer}. We also report computational cost for all compared methods, including FLOPs, peak memory, parameter count, and inference time. 

As shown in Table~\ref{tab:supp_baselines}, MedGSSR achieves the best PSNR and SSIM, while requiring the fewest FLOPs and the shortest inference time. Although ArSSR obtains the lowest LPIPS in this setting, it shows lower PSNR/SSIM and higher computational cost than MedGSSR. The two transformer-based baselines, MTVNet and SuperFormer, achieve competitive reconstruction quality but require substantially higher FLOPs and inference time. These results further demonstrate that MedGSSR offers a favorable accuracy-efficiency trade-off under domain shift.

\begin{table*}[h]
\vspace{-0.2cm}
\renewcommand{\arraystretch}{1.1}
\footnotesize
\centering
\caption{\textbf{Extended quantitative and computational comparison under cross-domain $4\times$ 3D SR on the HCP dataset.} In addition to the baselines reported in the main paper, MTVNet and SuperFormer are included as transformer-based baselines. Computational cost is reported for all methods. Best and second-best results are bolded and underlined, respectively.}
\vspace{-0.2cm}
\resizebox{\textwidth}{!}{%
\begin{tabular}{l@{\hspace{0.4cm}}c@{\hspace{0.35cm}}c@{\hspace{0.35cm}}ccccc}
\toprule
\textbf{Methods}  & \textbf{PSNR $\uparrow$} & \textbf{SSIM $\uparrow$} & \textbf{LPIPS $\downarrow$} & \textbf{\makecell{FLOPs\\(G)}} & \textbf{\makecell{Peak \\Mem. (GB)}} & \textbf{\makecell{Params \\(M)}} & \textbf{\makecell{Inference\\ Time (ms)}} \\
\toprule
MTVNet & $\underline{34.16{\pm}2.86}$ & $0.9246{\pm}0.0267$ & $0.1368{\pm}0.0416$ & $1359.50$ & $1.75$ & $22.61$ & $1083.18$ \\
SuperFormer & $33.87{\pm}3.02$ & $0.9142{\pm}0.0273$ & $0.1446{\pm}0.0435$ & $1888.20$ & $2.06$ & $19.66$ & $1592.91$ \\
CuNeRF & $33.12{\pm}3.56$ & $\underline{0.9382{\pm}0.0365}$ & $0.1587{\pm}0.0557$ & $2061.58$ & $\mathbf{0.12}$ & $\mathbf{0.98}$ & $259.76$ \\
ArSSR & $31.87{\pm}3.37$ & $0.9220{\pm}0.0324$ & $\mathbf{0.1141{\pm}0.0513}$ & $\underline{955.15}$ & $8.15$ & $\underline{6.28}$ & $\underline{105.98}$ \\
MedGSSR(Ours) & $\mathbf{35.84{\pm}2.73}$ & $\mathbf{0.9533{\pm}0.0224}$ & $\underline{0.1253{\pm}0.0378}$ & $\mathbf{385.87}$ & $\underline{1.29}$ & $18.13$ & $\mathbf{56.52}$ \\
\bottomrule
\end{tabular}
}
\label{tab:supp_baselines}
\vspace{-0.6cm}
\end{table*}

\section{Additional Visualization of Downstream Segmentation on Super-Resolved MRI Volumes}
To complement the downstream segmentation results reported in the main paper, we provide additional visual comparisons on the unseen HCP dataset using the same segmentation setting as described in the main paper. In particular, while the main paper reports quantitative segmentation results and includes one representative visualization for the 4$\times$ case, here we present more comprehensive visual results across axial, coronal, and sagittal views to further illustrate the impact of super-resolution quality on the downstream segmentation performance.

\begin{figure*}[t]
\centering
\includegraphics[width=\textwidth]{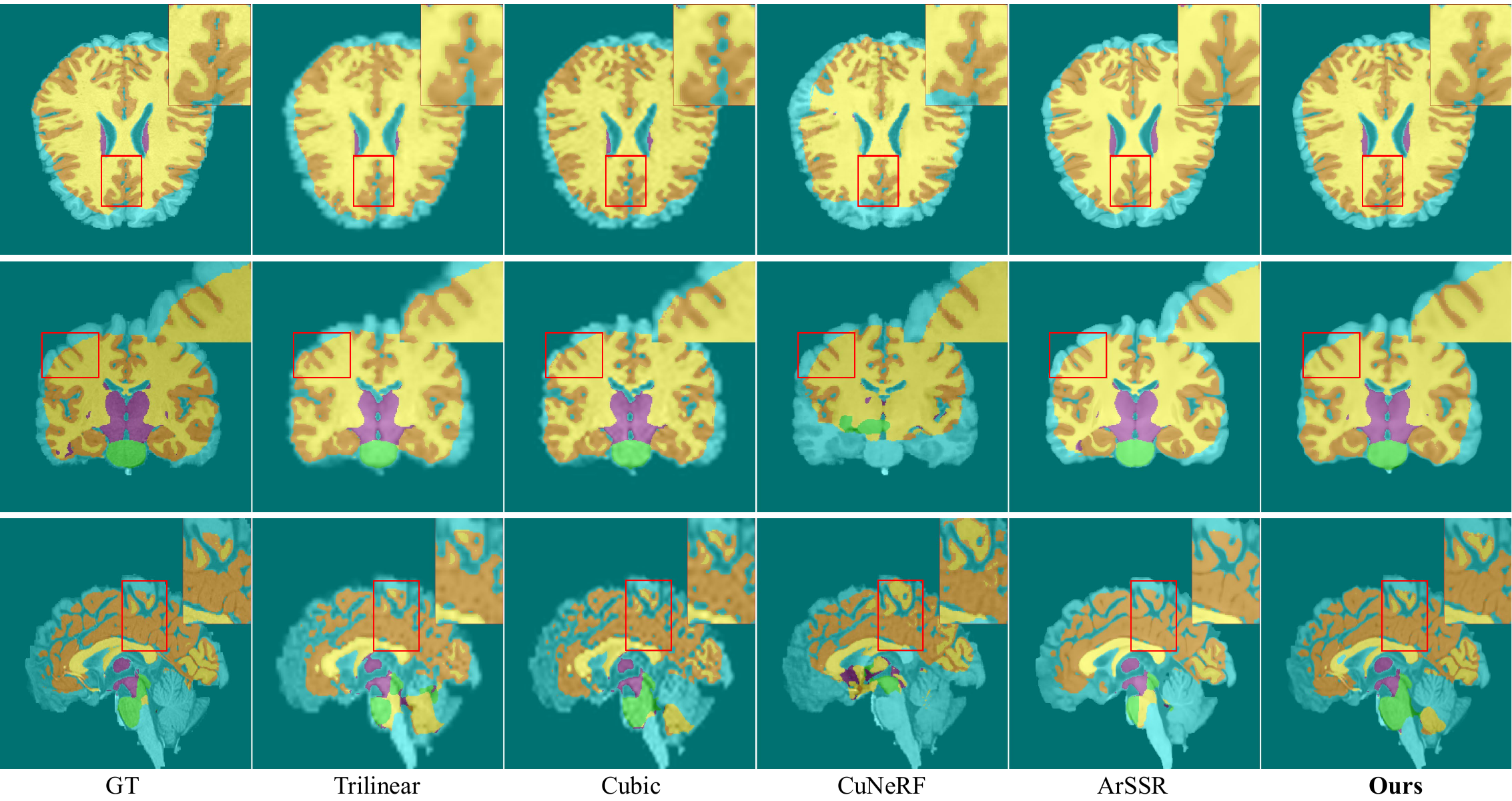}
\caption{Additional visual comparisons for downstream brain tissue segmentation under cross-domain on the unseen HCP dataset based on 4$\times$ super-resolved MRI volumes. Results are shown across axial, coronal, and sagittal views, with zoomed-in regions for detailed inspection. Compared with competing methods, MedGSSR yields segmentation masks that are more consistent with the ground truth, with clearer boundaries and more faithful anatomical structures.}
\label{fig:supp_seg4}
\end{figure*}

Figure~\ref{fig:supp_seg4} shows additional visual comparisons for downstream brain tissue segmentation based on 4$\times$ super-resolved MRI volumes. Consistent with the quantitative and visual results in the main paper, the segmentation masks obtained from MedGSSR are more faithful to the ground truth, with clearer structural boundaries and better preservation of fine anatomical regions. In contrast, the competing super-resolution methods introduce blurring, boundary leakage, or structural distortion, which propagate to the downstream segmentation results and lead to visibly less accurate tissue delineation.

\begin{figure*}[htbp]
\centering
\includegraphics[width=\textwidth]{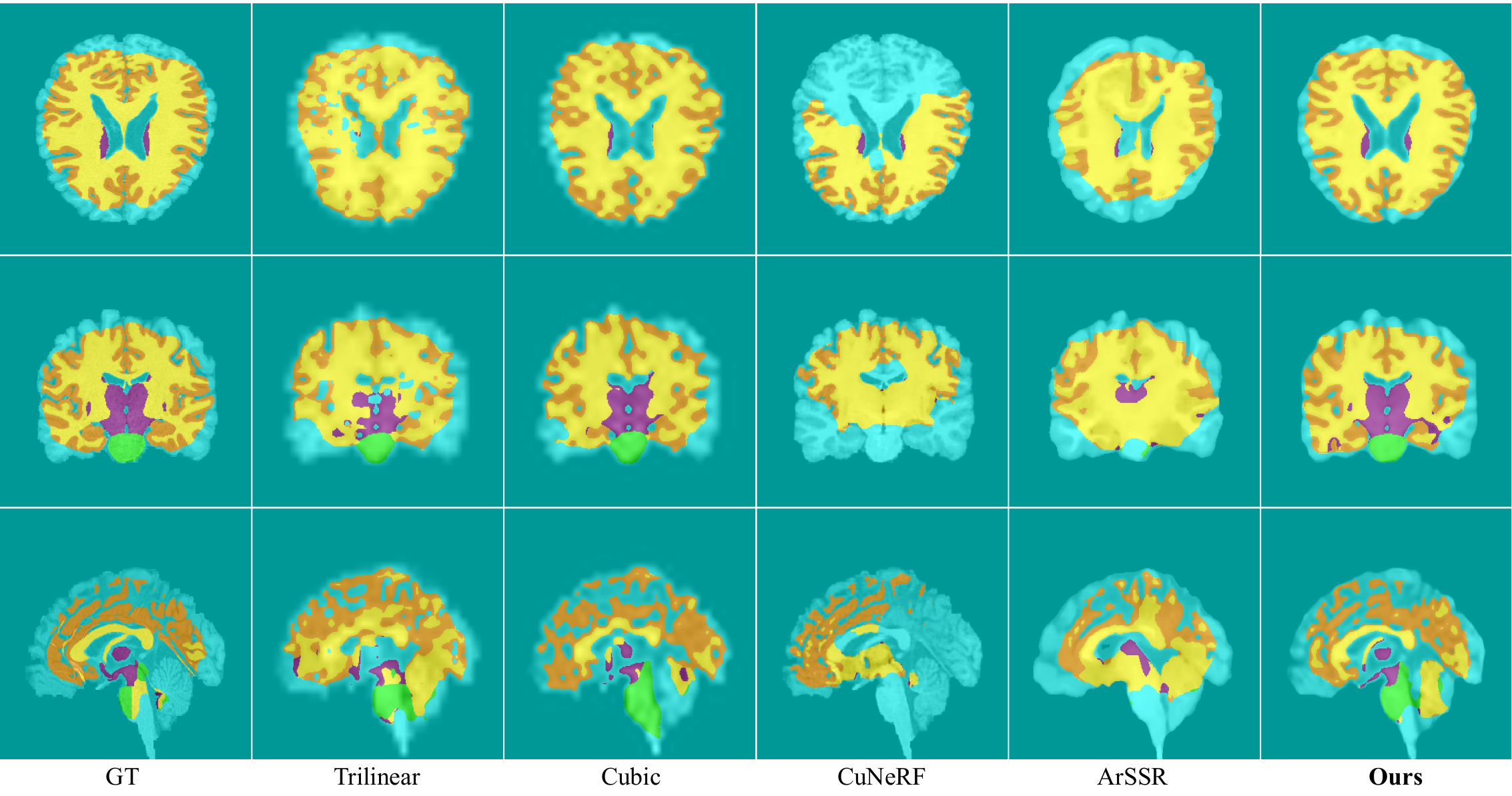}
\caption{Additional visual comparisons for downstream brain tissue segmentation under cross-domain on the unseen HCP dataset based on 8$\times$ super-resolved MRI volumes. Results are shown across axial, coronal, and sagittal views. This more challenging setting serves as an additional stress test beyond the main experiments. As the upsampling factor increases, the competing methods exhibit substantial degradation in structural consistency, whereas MedGSSR still preserves more faithful anatomical regions and produces segmentation masks that remain noticeably closer to the ground truth.}
\label{fig:supp_seg}
\end{figure*}

Figure~\ref{fig:supp_seg} further presents a more challenging 8$\times$ setting as an additional stress test beyond the main experiments. As the resolution gap becomes substantially larger, the quality of the super-resolved input becomes even more critical for reliable downstream analysis. Under this challenging setting, the baseline methods exhibit much stronger degradation in the predicted segmentation masks, including boundary confusion, region bleeding, and loss of structural consistency. By comparison, MedGSSR still preserves more coherent anatomical structures and produces segmentation results that remain substantially closer to the ground truth across all three views. These visual results further support that the proposed method preserves task-relevant structural information effectively, even under large upsampling factors and cross-dataset generalization.

\begin{table}[!t]
\centering
\caption{Additional CT ablation on the MELA dataset under the intra-domain $4\times$ 3D SR setting. The results show trends consistent with the MRI ablations and evaluate sub-voxel decomposition, Gaussian truncation radius, and reduced training data. Best results are highlighted in bold.}
\label{tab:ablation_ct}
{
\renewcommand{\arraystretch}{0.85}
\begin{tabular*}{\linewidth}{@{\extracolsep{\fill}}lccc}
\toprule
Variants & PSNR $\uparrow$ & SSIM $\uparrow$ & LPIPS $\downarrow$ \\
\midrule
Ours & $\mathbf{37.31}$ & $\mathbf{0.9362}$ & $\mathbf{0.1472}$ \\
Ours w/ $m=1$ & $36.42$ & $0.9234$ & $0.1564$ \\
Ours w/ $1\sigma$ truncation & $33.47$ & $0.8971$ & $0.1827$ \\
Ours w/ 75\% training data & $37.04$ & $0.9338$ & $0.1502$ \\
\bottomrule
\end{tabular*}
}
\end{table}

\section{Additional CT Ablation Study}
To further examine whether the design observations on MRI generalize to another imaging modality, we conduct additional ablations on the MELA CT dataset under the intra-domain $4\times$ 3D SR setting. As shown in Table~\ref{tab:ablation_ct}, the CT results exhibit trends consistent with the MRI ablations: reducing model capacity or narrowing the Gaussian support consistently degrades reconstruction quality, while moderate data reduction only causes a limited performance drop. Specifically, the full model achieves the best performance across all metrics. Reducing the sub-voxel count to $m=1$ decreases PSNR by 0.89~dB and worsens LPIPS, confirming the importance of sub-voxel Gaussian decomposition for representing fine CT structures. The $1\sigma$ truncation variant leads to the largest degradation, indicating that sufficient Gaussian support is also critical for CT reconstruction. In addition, the model trained with 75\% of the data remains close to the full-data model, demonstrating stable performance under reduced supervision.

\end{document}